\documentclass{optica-article}

\journal{opticajournal} 

\articletype{Research Article}

\usepackage{lineno}
\usepackage{physics}
\usepackage{cleveref}
\usepackage{hyperref}

\begin{document}

\title{Transfer of entanglement from nonlocal photon to non-Gaussian quantum states}

\author{Mikhail S. Podoshvedov\authormark{1,2} and Sergey A. Podoshvedov\authormark{1,2*}}

\address{\authormark{1}Laboratory of quantum information processing and quantum computing, laboratory of quantum engineering of light, South Ural State University (SUSU), Lenin Av. 76, Chelyabinsk, Russia\\
\authormark{2}Laboratory of quantum engineering of light, South Ural State University (SUSU), Lenin Av. 76, Chelyabinsk, Russia\\}

\email{\authormark{*}sapodo68@gmail.com} 


\begin{abstract*} 
Continuous variable (CV) entanglement refers to the type of entanglement of quantum wave-like systems that are described by continuous variables in an inherently infinite-dimensional space. It can become a crucial resource for quantum communication, sensing and computation. We propose the mechanism of transfer of quantum entanglement (TQE) from a nonlocal photon to two initially separate single-mode squeezed vacuum (SMSV) states. The nonlocal photon is the only original quantum resource from which entanglement is transferred to CV states of a certain parity in a deterministic manner without them directly interacting with each other. Measurement induced CV parity entanglement is tuned using initial squeezing and the beam splitter (BS) parameter allowing us to estimate the probability of transfer of maximum entanglement at sufficiently high brightness to be 0.2344 for initial SMSV states. If, instead of the original SMSV states, we use those from which one photon is initially subtracted, then the heralded technique can turn the maximum entanglement probabilistic transfer protocol into a nearly deterministic one, the probability of which is $> 0.98$. Such a perfect TQE from the nonlocal photon to a maximally parity-entangled CV state can be considered the most suitable for applications, since it preserves the trade-off between the probability of its implementation and brightness of the output non-Gaussian states. 

\end{abstract*}

\section{Introduction}
 Two quantum systems can be more strongly correlated than is possible in classical systems, allowing the generation of nonclassical statistics of measured outcomes. This phenomenon in quantum mechanics is known as entanglement \cite{1}\cite{2}. Regardless of the physical platform \cite{1}, the entanglement plays a key role in the implementation of the teleportation protocol \cite{3}\cite{4}, in quantum dense coding \cite{5}, in the ultra-precise estimate of unknown parameter \cite{6}\cite{7} and in quantum computing \cite{8}\cite{9}. The quantum network can be seen as a natural extension of the concept of entanglement. It consists of many quantum nodes processing and storing quantum information and quantum channels connecting them \cite{10}\cite{11}. Therefore, it is important to control and tune the entanglement that can be distributed over long distance, while maintaining coherence of the entangled states. 
\par Implementation of the optical quantum information has been traditionally based on use of the states with discrete variables (DV) and continuous variables \cite{12}\cite{13}. Single photon in two spatial modes \cite{14} and two photons in orthogonal polarizations \cite{15} are the examples of the DV approach. If amplitude and phase quadratures are used to encode quantum information in a quantum state, then we deal with CV approach to describe evolution of a quantum system \cite{16}\cite{17}\cite{18}\cite{19}. Two-mode entangled states can be quite routinely realized in optical systems in a deterministic manner using parametric down-conversion or, given the SMSV state \cite{20}, one can mix it with vacuum at the beam splitter to obtain the target CV state. The beam splitter acts as a Gaussian unitary operation, transforming the pure single-mode squeezed state into an entangled one, but without changing its Gaussian character. In practice, there are limits on the squeezing of the CV light \cite{21}\cite{22}, so that the contribution of the vacuum state dominates the output state, significantly reducing the average number of photons in the state. Another type of entangled CV states is realized in the presence of a nonclassical superposition of coherent states (SCSs) by passing it through a balanced BS. Despite the apparent simplicity of implementing the coherent entanglement, generating SCSs requires significant resources and effort \cite{23}\cite{24}\cite{25}\cite{26}\cite{27}. CV entanglement realized through the displacement operator has been studied in \cite{5}\cite{28}.    
\par Recently, significant progress has been made in combining both approaches, which implies the implementation of hybrid entangled states \cite{29}\cite{30}\cite{31}\cite{32}. Probabilistic preparation of the hybrid entangled state by heralded detection of a single photon has been demonstrated in \cite{31}, while deterministic generation of non-ideal hybrid entanglement is considered in \cite{32}. The hybrid entangled states have found wide application due to their ability to combine the advantages of DV and CV components \cite{33}\cite{34}\cite{35}\cite{36}. Next technological progress is associated with the reversible transformation of quantum states from one encoding to another and the implementation of quantum networks consisting of many nodes and channels connecting them. Entanglement swapping showing a capability to distribute the hybrid entanglement from DV-DV and DV-CV entanglement has been demonstrated in \cite{37}. Using optical hybrid entanglement, DV qubit can be converted into CV qubit \cite{38}. For practical use, entanglement generation must be deterministic or nearly deterministic, otherwise, when entanglement between two nodes is generated probabilistically, the probability distribution of the total entanglement decreases exponentially with scaling of the entangled state. This circumstance can place increased demands on the entanglement creation process, regardless of the physical platform used.
\par The goal of our work is to develop an efficient mechanism for generating CV entanglement from two initially separate CV states. The two original SMSV states are entangled by a nonlocal photon, so we can talk about the transfer of quantum entanglement from the DV state to the CV states. The term TQE will be used further when discussing the details of the mechanism for creating CV entanglement. The TQE resembles the standard quantum swapping protocol \cite{39}, but differs from it by reducing the number of initial resources (one entangled state instead of two) and by eliminating Bell state measurement (BSM), which is an integral part of optical quantum protocols with DV-DV and DV-CV encoding. Adding a nonlocal photon to the two SMCV states and simultaneously subtracting the photons by the genuine transition-edge sensors (TES detectors) \cite{40} in the two measurement channels allows them to be entangled in a deterministic manner but with a non-maximal degree of entanglement. Freely propagating CV states that have not been dynamically coupled in any way become entangled due to interactions with the initially prepared single photon entanglement. The output CV entanglement is maximized by optimizing the input parameters: squeezing of the SMSV state and the beam splitter parameter. The efficiency of the TQE from nonlocal photon to two SMSV is turned out to be limited as the generation of maximum CV entanglement is probabilistic. However, if we initially subtract the single photon from the SMSV states, the resulting TQE from the nonlocal photon to two odd CV states can be realized more efficiently. Thus, this protocol can be implemented in a nearly deterministic manner while maintaining a sufficiently large average number of photons \cite{41} in the CV states forming the output CV entanglement.

\section{TQE from nonlocal photon to initially separated SMSV states}

\begin{figure}[htbp]
\centering\includegraphics[width=0.5\textwidth]{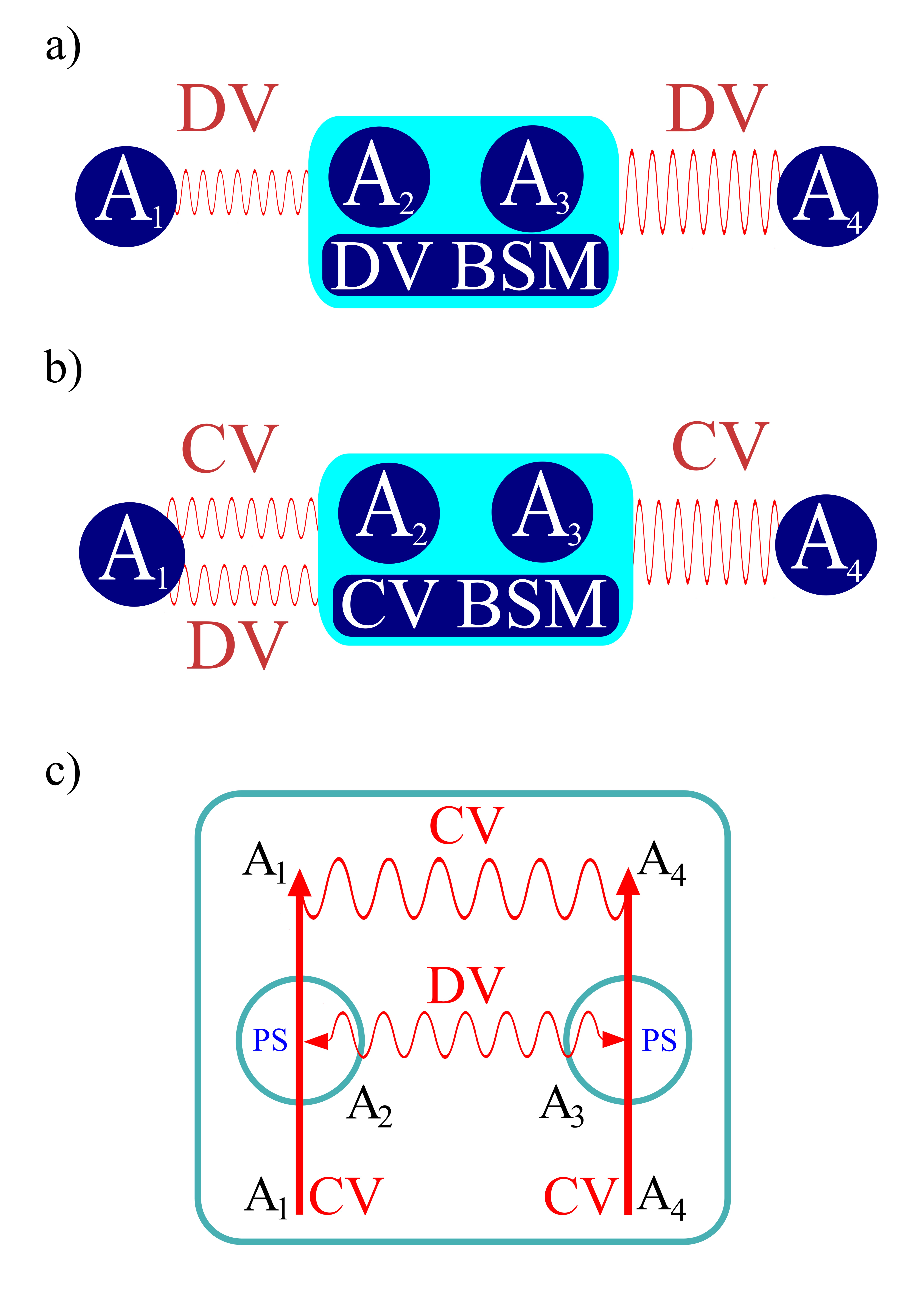}
\caption{(a-c). Comparison of (a) DV-DV ($A_1$-$A_2$ and $A_3$-$A_4$ nodes) and (b) CV/DV-CV ($A_1$-$A_2$ and $A_3$-$A_4$ nodes) swapping via either (a) DV BSM or (b) CV BSM and (c) TQE from DV quantum resource to initially separated CV states. The red vertical arrows correspond to the separate states in the nodes $A_1$ and $A_4$ which become connected at the top. The notations DV and CV indicate the initial DV and final CV-CV entanglement and PS means photon subtraction.}
\label{Figure.1}
\end{figure}

Here we are going to develop an approach aimed at entangling two CV states of definite parity, so that they do not directly interact with each other. In contrast to DV-DV in \autoref{Figure.1}(a) or DV-CV in \autoref{Figure.1}(b) swapping, only single nonlocal photon as a carrier of initial entanglement is used as shown in \autoref{Figure.1}(c). The addition of a nonlocal photon followed by subtraction of photons from the SMSV states represents the TQE mechanism from the DV state to two initial Gaussian states, which upon interaction with the nonlocal photon are transformed into non-Gaussian states with preservation of their parity. 
\par To implement the TQE mechanism in practice a nonlocal balanced photon 

\begin{equation}\label{eq:1}
\ket{\xi} =
       \frac{\ket{01}_{34} + \ket{10}_{34}}{\sqrt{2}}.
\end{equation}

\!\!\!\!\!\!is used which propagates in two separate spatial modes 3 and 4. The maximally entangled state of the single photon is obtained when it impinges on the balanced BS. Two identical SMSV states 

\begin{equation}\label{eq:2}
\ket{SMSV(y)} =
\frac{1}{\sqrt{\cosh{s}}}\sum_{n=0}^{\infty} \frac{y^{n}}{\sqrt{(2n)!}}\frac{(2n)!}{n!}\ket{2n}  
\end{equation}

\!\!\!\!\!\!distributed in modes 1 and 2 are selected as input. To characterize the SMSV state, squeezing parameter $y = {(\tanh{s})/2}$
is used determined through the squeezing amplitude $s>0$, which imposes a limitation $0\leq{y}\leq{0.5}$ on it. In addition, the squeezing $S=-10 \log{(\exp{(-2s)})}$  dB in decibels and the average number of photons $\langle n_{SMSV} \rangle$ =$\sinh^{2} s$ can be used to characterize the original SMSV states. In the case under consideration, the identity of the two SMSV states means that their squeezing $S_1$ and $S_2$ are equal to each other, i.e. $S_1$=$S_2$=$S$.

\begin{figure}[htbp]
\centering\includegraphics[width=0.5\textwidth]{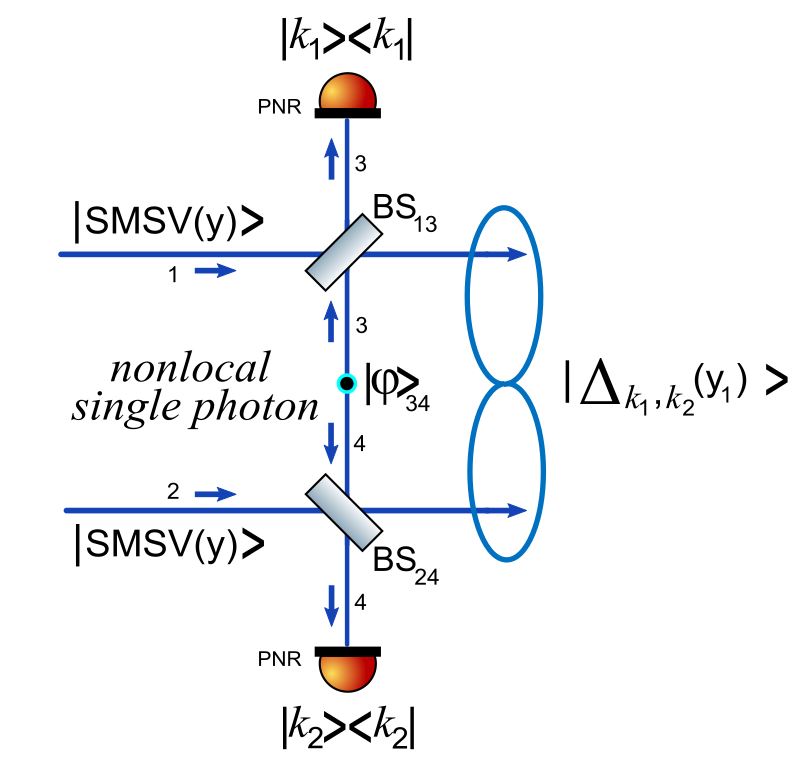}
\caption{A realistic scheme used for the TQE from a nonlocal photon to two input SMSV states that are distant from each other and do not directly interact with each other. Each of the SMSV states interacts with the nonlocal photon on two beam splitters, with subsequent registration of coinciding photons in measuring modes by PNR detectors. Perfect TQE with maximum output CV entanglement is only possible under certain conditions. To turn the perfect TQE protocol into a nearly deterministic one, it is necessary to use CV states in equation \eqref{eq:12} instead of SMSV states.}
\label{Figure.2}
\end{figure}

\par In practice, the TQE can be realized as shown in \autoref{Figure.2} by mixing two SMSV states with a nonlocal photon at two identical BSs. The identity of two beam splitters means that their real transmittances $t_1>0$,$\,\,t_2>0$ and reflectances $r_1>0$,$\,\,r_2>0$ are equal to each other, i.e. $t_1$=$t_2$=$t$ and $r_1$=$r_2$=$r$. One of the $BSs$ in \autoref{Figure.2} with subscripts 13 mixes the SMSV state in the first mode with mode 3 of the nonlocal photon. Accordingly, the second $SMSV$ state in mode 2 interacts with mode 4 of the nonlocal photon on the beam splitter $BS_24$. As follows from figure 2, modes 3 and 4 are measuring, that is, those in which photon number resolving (PNR) detectors are located. The relevant technical details of the TQE are presented in the supplementary material.
As follows from the supplementary material, the joint registration of the measurement outcomes $(k_1k_2)$ in the measured modes in figure 2 allows for one to conditionally realize the entangled state

\begin{equation}\label{eq:3}
    \ket{\triangle_{k_1k_2}(y)} =
    \frac{1}{\sqrt{N_{k_1k_2}}}\Bigl(
    \ket{\Psi_{k_1}^{(0)}(y_1,B)}_1
    \ket{\Psi_{k_2}^{(1)}(y_1,B)}_2 +
    b_{k1k2}\ket{\Psi_{k_1}^{(1)}(y_1,B)}_1
    \ket{\Psi_{k_2}^{(0)}(y_1,B)}_2\Bigl)
\end{equation}

\!\!\!\!\!\!where $N_{k_1k_2}=1+b_{k_1 k_2}^2$ is the normalization coefficient. Here, the $CV$ non-Gaussian states of a certain parity $\ket{\Psi_{k}^{(0)}(y)}$ and $\ket{\Psi_{k}^{(1)}(y)}$ are used which depend on the value of the squeezing parameter reduced by the value of the transmission coefficient $T={t}^2$. In terms of the beam splitter parameter $B=r^{2}⁄t^{2}$ , the reduced squeezing parameter can be rewritten as $y_{1}=yt^{2}=y⁄(1+B)$ since $T=1⁄(1+B)$, while the reflection coefficient of the BS is $R=r^{2}=B⁄(1+B)$. Note that the subscript $k$ in the $CV$ states corresponds to the number of photons subtracted, while the superscript indicates the number of photons added, either $0$ or $1$. The $CV$ states with the same number of subtracted photons $k$ but with superscripts $0$ and $1$ differ from each other in parity so that their overlap is zero ${\bra{\Psi_{k}^{(0)}}\ket{\Psi_{k}^{(1)}}}$, that is, they are orthogonal to each other. The form of the measurement induced CV states of definite parity is presented in the supplementary material. Note that the generation of the entangled state in Eq.\eqref{eq:3} occurs when two PNR detectors distinguish the number of incoming photons, each in its own measuring channel \cite{27}.
Internal multiplier $b_{k_1 k_2}$ of the CV entangled state arises from the fact that the amplitudes of the entangled hybrid state $c_k^{(0)}(y_1,B)$ and $c_k^{(1)}(y_1,B)$ emerging from the interaction of the states with the nonlocal photon on two identical $BSs$ as in \autoref{Figure.2} are not equal to each other. It is given by  

\begin{equation}\label{eq:4}
    b_{k_1k_2}(y_1,B) =
    \begin{cases}
            \,\,\,\,\,\,\,\,\,\,\,\,\,\,\,\,\,\,\,\,\,\,\,\,\,\,\,\,\,1, \,\,\text{if}\,\,\, k_1 =k_2 = 0 \\[3ex]
			-\sqrt{B}\frac{\sqrt{y_1B}}{k_2}\sqrt{\frac{G_0^{(1)}(y_1,B)Z^{(k_2)}(y_1)}{G_{k_2}^{(1)}(y_1,B)Z(y_1)}}, & 
            \,\text{if } k_1 = 0,  k_2 > 0 \\[3ex]
			-\frac{1}{\sqrt{B}}\frac{k_1}{\sqrt{y_1B}}\sqrt{\frac{G_{k_1}^{(1)}(y_1,B)Z(y_1)}{G_0^{(1)}(y_1,B)Z^{(k_1)}(y_1)}}, &\,\,\text{if } k_2 = 0,k_1 > 0 \\[3ex]
            \frac{k_1}{k_2}\sqrt{\frac{G_{k_1}^{(1)}(y_1,B)Z^{k_2}(y_1)}{G_{k_2}^{(1)}(y_1,B)Z^{k_1}(y_1)}}, & \,\,\,\text{if }
            k_1 > 0, k_2 > 0
	\end{cases}
\end{equation}

The details of the derivation of the amplitudes $c_k^{(0)}(y_1,B)$ and $c_k^{(1)}(y_1,B)$ as well as the multiplier in equation \eqref{eq:4} are presented in the supplementary material. 
The factor $b_{k_1k_2}$ can rightly be called distorting in the sense that it reduces the entanglement of the generated state, which can be expressed, for example, through negativity $N$ having all required properties of the entanglement measure \cite{30}. Indeed, regardless of the values of experimental input parameter $S$ and $B$, the generated entangled states in equation \eqref{eq:3} can be considered in a four-dimensional Hilbert space because the $CV$ states with different superscripts in one mode differ in their parity. This allows us to derive expression for the negativity $\mathcal{N}_{k_1k_2} =2b_{k_1k_2}⁄N_{k_1k_2}$ , where the maximally entangled state has $\mathcal{N}_{k_1k_2}=1$ that is possible only in the case of $b_{k_1k_2}=1$ which automatically entails that the normalization factor becomes equal to two, that is, $N_{k_1k_2}=2$. Here we define the negativity of the full system as $\mathcal{N}_{k_1k_2}=‖\rho_{k_1 k_2}^{T_1}‖-1$ for the generated state $\rho_{k_1 k_2}$ following from equation \eqref{eq:3}, where $\rho_{k_1 k_2}^{T_1}$ is the partial transpose of $\rho_{k_1 k_2}$ with respect to the first subsystem mode 1 and $‖\rho_{k_1 k_2}^{T_1}‖$ is the trace norm of the density matrix. We can also use the logarithmic negativity $E_\mathcal{N}(\rho_{k_1 k_2})=\log_2(N_{k_1k_2}+1)$. 
as a quantitative measure of $CV$ entanglement which is computable with known value of $\mathcal{N}_{k_1 k_2}$.          
\par As follows from the definition of the amplitude-distortion coefficient in equation \eqref{eq:4}, it becomes equal to 1 in the case of equality of the measurement outcomes $k_1=k_2=k$, i.e. $b_{kk}=1$ regardless of the parameter values $S$ and $B$. The amplitude distorting factors with rearranged subscript $k_1\leftrightarrow	k_2$ are related to each other as reciprocal values, that is, $b_{k_1 k_2}=1⁄b_{k_2 k_1}$  what is used next. The values of the function $b_{k_1 k_2}$ never take zero values $(b_{k_1k_2}\neq0)$, which entails a non-zero value of the negativity $\mathcal{N}_{k_1 k_2}\neq0$ and then we can talk about the deterministic implementation of the $TQE$ from a nonolocal photon to non-Gaussian states not interacting with each other. Here, by deterministic implementation we mean that $TQE$ is possible for any measurement outcomes. Nevertheless, in most cases of the measurement outcomes the $CV$ entanglement is not maximal. In what follows, we will call the $TQE$ with maximum output $CV$ entanglement perfect. In general, the problem can be formulated as a search for conditions for the perfect implementation of the $TQE$.
\par The success probability in registering coincidences and thus in the conditional formation of measurement induced $CV$ entanglement follows from the amplitudes $c_k^{(0)}(y_1,B)$ and $c_k^{(1)}(y_1,B)$ of the entangled state at the output of two identical $BSs$

\begin{equation}\label{eq:5}
    P_{k_1k_2}(y_1,B) = \frac{N_{k1k2}(1 - 4y_{1}^{2}(1 + B)^2)}{2(2+B)}
    \begin{cases}
            BZ(y_1)G_0^{(1)}(y_1,B), \,\,\text{if}\,\,\, k_1 =k_2 = 0 \\[3ex]
			B\frac{(y_1B)^{k_1}}{k_1!}Z^{(k_1)}(y_1)G_0^{(1)}(y_1,B), & 
            \!\!\!\!\!\!\!\!\!\!\!\!\!\!\!\!\!\!\text{if } k_1>0, k_2 = 0 \\[3ex]
            
			\frac{(y_1B)^{(k_1+k_2-1)}}{k_1!k_2!} k ^2_2 Z ^{(k_1)}(y_1)G_{k_2}^{(1)}(y_1,B), & \!\!\!\!\text{if } k_1 \ge   0,k_2 > 0 
	\end{cases}
\end{equation}

which are present in the supplementary material. Note that the multiplier $N_{k_1 k_2}⁄2$ can act as an amplifying factor if $b_{k_1k_2}>1$. In the general case, the probabilities that differ by the permutation of the subscripts $k_1$ and $k_2$ are equal to each other, that is, $P_{k_1k_2}(y_1,B)=P_{k_2k_1}(y_1,B)$, which can be directly verified from the definition of the normalization and amplitude distortion factors. From a physical point of view, this can be interpreted as a manifestation of symmetry with respect to the sources of states and the optical elements used in \autoref{Figure.2}.
\par The probability distribution can be divided into two parts: the same measurement outcomes, i.e. those for which $k_1=k_2=k$ including $k=0$, and the outcomes with different numbers of measured photons, i.e. when $k_1\neq k_2$. This allows us to immediately identify that part of the distribution with $k_1=k_2=k$ that guarantees the generation of the maximum entangled state with the probability

\begin{equation}\label{eq:6}
    P_{k_1 = k_2} = P_{00} + \sum_{k=1}^{\infty} P_{kk}(y_1,B),
\end{equation}

where

\begin{equation}\label{eq:7}
    P_{00}(y_1,B) = \frac{1 - 4y_1^2(1+B)^2}{1+B}BZ(y_1)G_0^{(1)}(y_1,B) = \frac{(1 - 4y_1^2(1+B)^2)BZ^4(y_1)}{1+B}
\end{equation}

\begin{equation}\label{eq:8}
    P_{kk}(y_1,B) = \frac{1 - 4y_1^2(1+B)^2}{1+B} \frac{(y_1B)^{(2k-1)}}{(k!)^2} k ^2Z^{(k)}(y_1)G_k^{(1)}(y_1,B)
\end{equation}


\!\!\!\!\!\!since $N_{k_1k_2}=2$ in the case of $b_{kk}=1$. When deriving the expression for no click - no click probability $P_{00}$ we use $G_{0}^{(1)}(y_1,B)=Z^{3}(y_1)$ (formula (S11) in the supplementary material) and identity $1 / \cosh^2 s = 1 - 4y_{1}^{2}(1+B)^{2}$ for all probabilities of the measurement outcomes.

\par Naturally, it is worth starting the analysis with probabilities $P_{k_1=k_2}$ as the measured outcomes $k_1=k_2=k$ provide the maximal $CV$ entanglement. The probabilities depend on the parameter $B$, which allows us to optimize them, that is, find those optimizing values $B_{k_1=k_2}$ at which the optimized probability $P_{k_1=k_2}$ takes on a maximum value. In \autoref{Figure.2}  we show the dependence of the optimized over parameter $B$ sum of the probabilities $P_{kk}$ involving vacuum contribution and first four photonic states designated as $P_{0-4}=P_{00}+P_{11}+P_{22}+P_{33}+P_{44}$ and one without vacuum contribution $P_{1-4}=P_{11}+P_{22}+P_{33}+P_{44}$ on the values of the initial squeezing $S$. We use subscripts $0-4$ and $1-4$ for the probabilities in \autoref{Figure.2} as they include a limited number of terms which distinguishes them from complete probability $P_{k_1=k_2}$ in Eq. \eqref{eq:6} which involves all the measured outcomes with $k_1=k_2=k$. Choosing only the first four measured outcomes makes sense since they contribute the most to the overall probability. The optimizing values of $B_{0-4}$ and $B_{1-4}$ defining $P_{0-4}$ and $P_{1-4}$ are shown in \autoref{Figure.3(a,b)}(a,b) in dependency on the squeezing parameter $S$.

\begin{figure}[htbp]
\centering\includegraphics[width=0.9\textwidth]{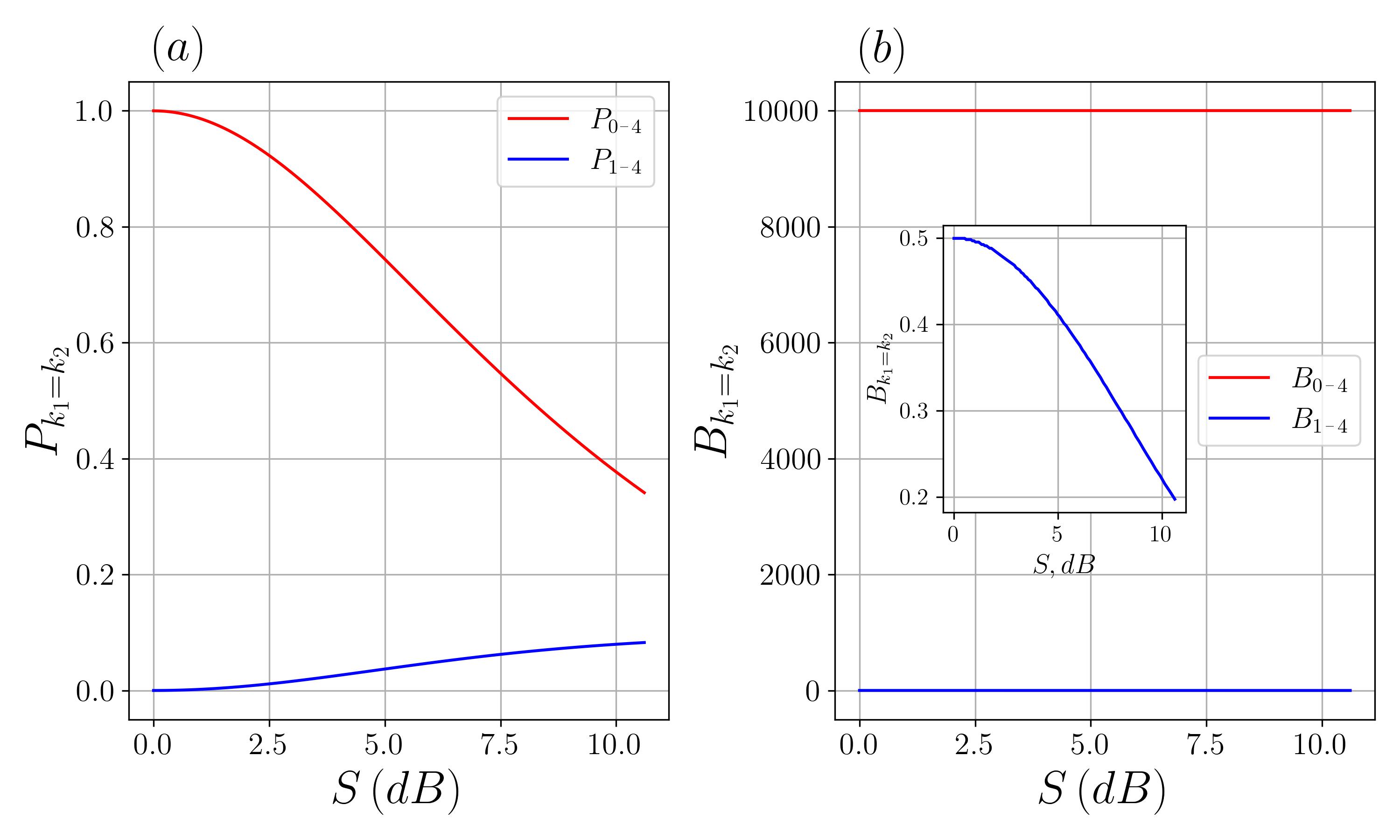}
\caption{(a,b) (a) Graphical dependencies of the success probabilities $P_{k_1=k_2}$ for the perfect $TQE$ obtained by optimization by parameter $B=B_{k_1=k_2}$on initial squeezing $S\,\,(dB)$ both with (curve $P_{0-4}$) and without (curve $P_{1-4}$) no click contribution, where subscripts $0-4$ and $1-4$ means taking into account the results of measurements $k=0,1,2,3,4$ and $k=1,2,3,4$, respectively. The optimizing values in (b) both $B_{0-4}$ and $B_{1-4}$ appear as two parallel horizontal lines due to the fact that $HRBSs$ with $B\gg1$ are required when the vacuum contribution is taken into account. In reality, the $B_{1-4}$ has a form as shown in the insert, which indicates the use of the beam splitters with larger transmissivity $T>R$.}
\label{Figure.3(a,b)}
\end{figure}

\par As can be seen from \autoref{Figure.3(a,b)}, the probabilities $P_{0-4}$ with and $P_{1-4}$ without the vacuum contribution behave in exactly the opposite way. If the vacuum contribution is included in calculation, then the probability of the maximum entanglement starts from a value close to 1 at $S\ll1$, i.e. $P_{0-4}\approx1$ with a smooth drop with increasing the initial squeezing. The probability without vacuum contribution takes the values close to zero, i.e. $P_{1-4}\approx0$, in the vicinity of $S\approx0$ and then it begins to grow with increasing $S$. The behavior of the optimizing values $B_{k_1=k_2}$ in \autoref{Figure.3(a,b)} depends on whether we take into account vacuum outcome $B_{0-4}$ or not $B_{1-4}$. If the vacuum contribution is taken into account then highly reflective beam splitters $(HRBSs)$ must be used since the value $B_{0-4}\gg1$ and, as consequence, $R\gg T$. Indeed, it follows from the analytical expression \eqref{eq:7} that the probability of no click – no click event turns into $P_{00}\approx B⁄(1+B)\rightarrow1$ with B growing in the vicinity of $S\approx0$ since $Z^4(y_1)\rightarrow1$ with $S\rightarrow0$. The optimizing parameter $B_{1-4}$ takes values less than 1 decreasing with increasing $S$, which corresponds to the use of the $BSs$ with $T> R$, where the case of $B=1$ corresponds to the balanced $BS$ with $T=R=0.5$. Since $B_{0-4}$ takes a very large and constant value, the values of the $B_{1-4}$ look like a straight line in the \autoref{Figure.3(a,b)} in the scale used for $B_{0-4}$. But if one reduces the scale of the change of B, then as shown in the insert, $B_{1-4}$ already takes the form of a descending curve starting from approximately $B_{1-4}\approx0.5$. 
\par Nearly unit probability of success $P_{0-4}\approx1$ for $S\ll1$ $dB$ is possible since $HRBSs$ largely redirects light energy into the measuring modes, thereby significantly reducing the mean number of photons in the $CV$ states forming the entangled state in Eq. \eqref{eq:3}, which is reflected in the parameter $y_1$ decreasing to almost zero, bringing the final state closer to a nonlocal photon. Indeed, taking into account the numerical data in \autoref{Figure.3(a,b)}, we have $y_1=y⁄(1+B_{0-4})\ll1$. As presented in the supplementary material and in \cite{41}, the average number of photons strongly depends on the parameter $y_1$, so the brightness of such $CV$ states significantly drops in the case of $B=B_{0-4}\gg1$. As for the measurement of vacuum fluctuations, a small increase in temperature occurs when the vacuum fluctuation interacts with the $TES$. $TES$ detectors have a high sensitivity to temperature changes, which allows them to detect very small thermal fluctuations caused by vacuum fluctuations \cite{40}. Although the perfect $TQE$ with the inclusion of the "no-click - no-click" event is implemented nearly deterministically, the significant reduction in the brightness of the output $CV$ entanglement due to the use of $HRBSs$ reduces the practical significance of such an implementation. Thus, the $TQE$ from nonlocal photon to two separated $SMSV$ states based on the $HRBSs$ cannot be considered significant.
\par On the contrary, as shown in the inset of \autoref{Figure.3(a,b)}, using the $BSs$ with $B_{CV}=B_{1-4}=0.221$ and $SMSV$ states with $S=10.6 \,\,dB$ allows $TQE$ to be realized without the contribution of vacuum measurement with the maximum success probability $P_{CV}=P_{1-4}=0.0827$. Here and below we use the subscript $CV$ for $P$ and $B$ indicating the parameter estimates of the perfect $TQE$, i.e., with maximal output entanglement. In the case, the squeezing parameter $y$ of the measurement induced $CV$ states decreases only by $1+B_{CV}$ times that is, it becomes $y_1=y⁄(1+B_{CV)}$, which is quite comparable with the initial $y$ as $B_{CV}<1$. This value of $y_1$ may even be sufficient to provide the mean number of photons greater than in the initial $SMSV$ states.     
\par
\begin{figure}[htbp]
\centering\includegraphics[width=\textwidth]{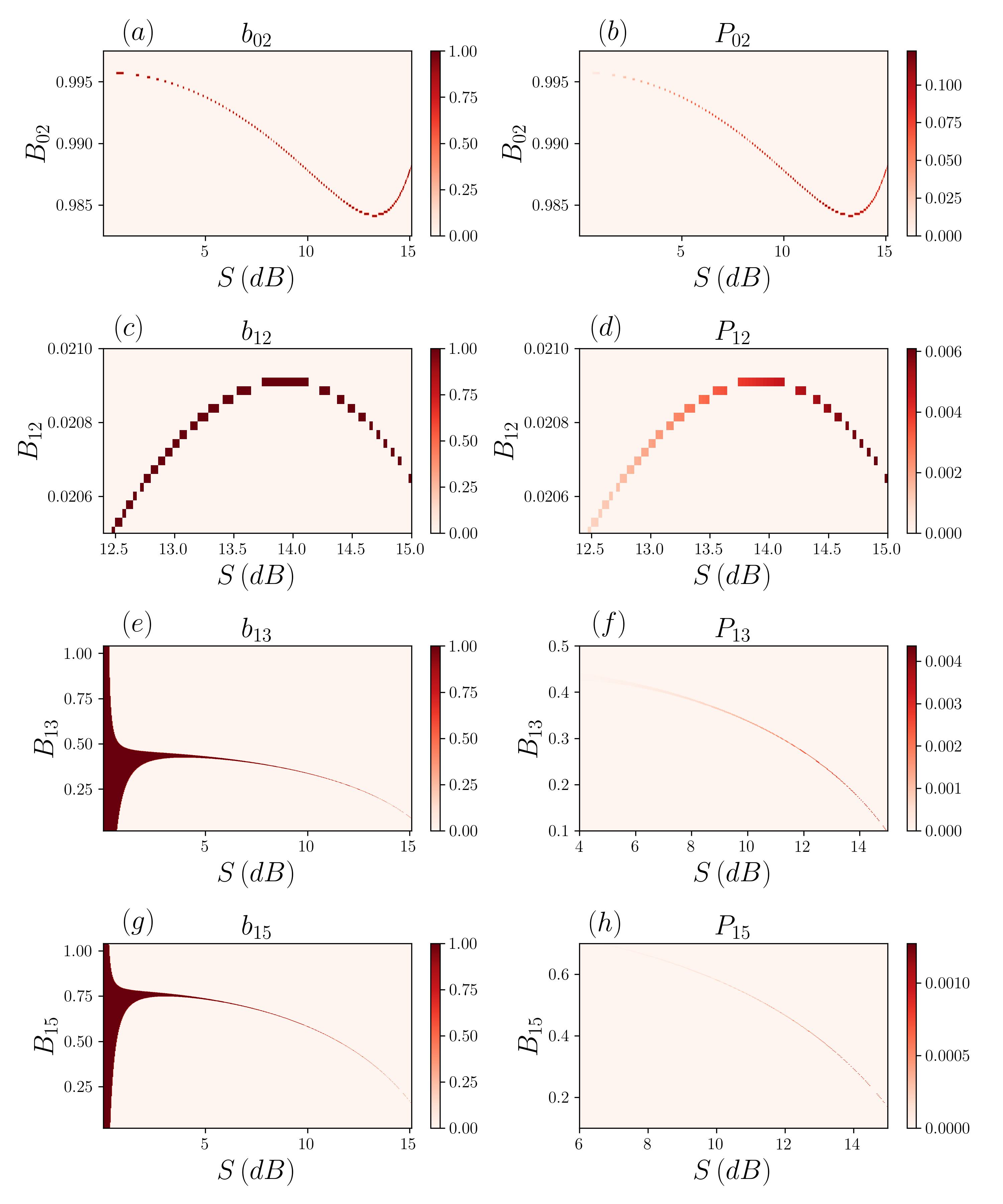}
\caption{(a-h). Contour graphs of the amplitude distortion factor $b_{ij}$ (on the left side, that is in a), c), e) and g)) and success probabilities $P_{ij}$ of generation of the $CV$ entanglement with help of a nonlocal photon (on the right side, that is, the graphs in (b), (d), (f) and (h)) in dependency on the squeezing S and BS parameter $B_{ij}$, where $i$ and $j$ are the number of measured photons. Visually, the absence of the amplitude distorting factor, i.e. $b_{ij}=1$ can be accurately determined by the shades of red: the darker the red, the closer $b_{ij}$ is to $1$. Corresponding probabilities $P_{ij}$ can also be estimated by the shades of red using a graduated color bar. The dark red curves correspond to the maximum values of $P_{ij}$. There is a correlation between the $b_{ij}=1$ curves and the maximum values of $P_{ij}$. The range of values of $S$ and $B$ is chosen to ensure the condition  $b_{ij}\le 1$. }
\label{Figure.4(a-h)}
\end{figure}

 A possible way to increase the success probability of generating the maximum $CV$ entanglement is to find values of $S$ and $B$ such that the amplitude-distortion factor $b_{k_1 k_2 }$ in Eq. \eqref{eq:4} is equal to one, i.e. $b_{k_1k_2}=1$ with $k_1\neq k_2$. Numerical search shows that such parameter values cannot be found for all $k_1$ and $k_2$. In particular, the amplitude distorting parameter when there is no click in the first $PNR$ detector and a single photon is registered in the second $PNR$ detector is not equal to one $b_{01}\neq 1$ in the selected range of change of $S$ and $B$. The same applies to the multiplier $b_{10}$. So single photon measurement event does not guarantee perfect $TQE$ implementation. To clearly demonstrate the possibility of fulfilling the condition $b_{ij}=1$, where for brevity $i=k_1$ and $j=k_2$ is used, \autoref{Figure.3(a,b)} shows the contour dependencies of $b_{ij}$ (graphs on the left) on $S$ and $B_{ij}$. The curves $b_{ij}=1$ are highlighted in dark red. The right side of \autoref{Figure.4(a-h)} can be used to estimate the probabilities $P_{ij}$ (dark red curves of a certain thickness) corresponding to $S$ and $B_{ij}$, which provides the implementation of the condition $b_{ij}=1$.

\par Numerical simulation shows the possibility of performing $b_{02}=b_{20}=1$ in two-photon subtraction using a near balanced $BS$ as its $B_{02}\approx1$, but nevertheless $B_{02}<1$ as shown in \autoref{Figure.3(a,b)}. The corresponding probabilities of no click – two photons and two photons – no click events follow from distribution in Eq. \eqref{eq:5} and can be rewritten as

\begin{equation}\label{eq:9}
    P_{02}(y_1,B) = \frac{2y_1BZ(y_1)G_2^{(1)}(y_1,B)}{(1+B)\cosh^2 s}, \,\,\,\,\,
    P_{20}(y_1,B) = \frac{2y_1^2B^3(1+2y_1^2)Z^3(y_1)}{(1+B)\cosh^2 s}
\end{equation}

where the normalization coefficient $G_2^{(1)}(y_1,B)$ is presented in the supplementary material. They are equal $P_{02}=P_{20}$, in particular, for those $B_{02}$ and $S$ for which the condition $b_{02}=b_{20}=1$ is met. The numerical data in \autoref{Figure.3(a,b)} make it possible to estimate parameters for the $TQE$ with maximal output entanglement: $P_{CV}=2P_{02}=2\cdot0.1172=0.2344$ given that $S=13.303\,\,dB$ and $B_{CV}=B_{02}=0.9814$. 
\par Another possibility to obtain unit amplitude distortion factor $b_{12}=b_{21}=1$ is associated with measurement events: single photon – two photons and two photons – single photon as shown in \autoref{Figure.3(a,b)}. In contrast to the case of $b_{02}=b_{20}=1$, the absence of the amplitude distorting multiplier $b_{12}=b_{21}=1$ is already observed for the beam splitters that are quite transmitting $B_{12}\ll1$. As shown in \autoref{Figure.3(a,b)}, the corresponding probabilities $P_{12}$ are already significantly less than those given above. Since the parameter $B_{12}$ shifts towards small values $B_{12}\ll 1$, it is incompatible with $B_{02}$. This does not allow adding probabilities $P_{12}=P_{21}$ in \autoref{Figure.3(a,b)} to $P_{02}=P_{20}$ to increase the overall probability $P_{CV}$ of creating the maximum $CV$ entanglement.

\par Among the measuring events with the subtraction of four photons in both channels it is possible to find conditions $B=B_{13}$ and $S$ as in \autoref{Figure.4(a-h)} such that ensure elimination of the amplitude distorting factor $b_{13}$ and $b_{31}$, that is $b_{13}=b_{31}=1$. However, the conditions that ensure $b_{13}=b_{31}=1$ do not correspond to those $B=B_{02}$ and $S$ under which $b_{02}=b_{20}=1$ is observed which does not allow adding the probability $P_{13}=P_{31}$ in \autoref{Figure.4(a-h)} to the overall probability $P_{CV}$ in the case of choosing condition adapted for two-photon subtraction in \autoref{Figure.4(a-h)}(a) and \autoref{Figure.4(a-h)}(b). 
Also in \autoref{Figure.4(a-h)} the dependencies of $B_{15}$ and $S$ are present, that is, in the case of subtraction of 6 photons in two measurement channels, for which the condition $b_{15}=b_{51}=1$ is fulfilled. Note a significant decrease in the probability of the events $P_{15}=P_{51}$ as shown in \autoref{Figure.4(a-h)} when subtracting $6$ photons compared to the probabilities $P_{02}=P_{20}$ of subtracting two photons. The probabilities of $P_{15}$ and $P_{51}$ also cannot be added to the total probability $P_{CV}$ if $B_{02}$ and $S$ are selected as in \autoref{Figure.4(a-h)}.
Thus, the best strategy for $TQE$ from a nonlocal photon to two input $SMSV$ states is to use $B=B_{02}$ and $S$ shown in \autoref{Figure.4(a-h)} adjusted for no click – two photons and two photons – no click measurement, which maximize the probability of $P_{CV}$ while maintaining proper brightness level in output $CV$ entanglement. This measurement is turned out to be more efficient compared to coincidence counting with the same numbers of photons in both channels resulting with probability of $P_{1-4}$. Such a $TQE$ can be considered optimal but not perfect due to the insufficient probability of realizing the maximally entangled $CV$ state.
The above analysis is presented in an ideal scenario, ignoring the imperfections of both the measuring detectors and the photon losses during their propagation. Specifically, it is assumed that the quantum efficiency of the $PNR$ detectors is equal to $1$, that is, $\eta=1$. Here, we limit our consideration to the influence of non-unit quantum efficiency on the output characteristics of the generated $CV$ entanglement. Each of the $PNR$ detector with quantum efficiency $\eta$ can be modeled using the positive operator-valued measure $(POVM)$ formalism with $POVM$ elements 
$\lbrace\hat{\Pi}_{k},k=0,\ldots,\infty \rbrace$, so that the measurement in the two measurement modes can be described by their tensor product $\hat{\Pi}_{k_1}$ $\hat{\Pi}_{k_2}$. Using the general expression for the entangled hybrid state before measurement in auxiliary modes (formula $(S19)$ in supplementary material) and applying the $POVM$ formalism, one obtains expressions for both the fidelity of the output $CV$ entangled state

\begin{equation}\label{eq:10}
    Fid_{k_1k_2} = 1 - (1 - \eta)\frac{y_1B}{N_{k_1k_2}}R_{k_1k_2}
\end{equation}

where 

\[
    R_{k_1k_2} = \frac{N_{k_1+1\,k_2}Z^{(k_1+1)}(y_1)}{Z^{(k_1)}(y_1)} + \Biggl(\frac{k_2+1}{k_2}\Biggl)^2 \frac{N_{k_1\,k_2+1}G_{k_2+1}^{(1)}(y_1,B)}{G_{k_2}^{(1)}(y_1,B)}
\]

for $k_{2}>0$ and the probability of realizing the target entanglement

\begin{equation}\label{eq:11}
    P_{k_1k_2,\eta} = \eta^{k_1+k_2}P_{k_1k_2}\Biggl(1 + (1-\eta)\frac{y_1B}{N_{k_1k_2}}R_{k_1k_2}\Biggl)
\end{equation}

\!\!\!\!\!in the first order in the smallness parameter $1-\eta\ll 1$, where $P_{k_1 k_2}$ is given by the formula \eqref{eq:5}. 

\par These expressions can form the basis for analyzing the impact of the quantum efficiency of $PNR$ detectors on the fidelity and probability of the $TQE$ protocol in a non-ideal scenario if the quantum efficiency is close enough to $1$, that is, $\eta\approx1$ but $\eta<1$. If the quantum efficiency is significantly less than $1$, then the expressions for the fidelity and probability must be supplemented with terms proportional to $(1-\eta)^k$, where $k>1$. Consideration of the $TQE$ protocol, given the imperfection of the measuring equipment, requires separate consideration. Let us consider only the special events of no click – two photons and two photons – no click, probabilities of which in the ideal case are presented in equation \eqref{eq:9}. If we use highly efficient quantum detectors with quantum efficiency $\eta=0.98$ \cite{40}, then we can estimate the fidelity as $Fid_{k_1 k_2}=0.976$ and the probability remains approximately at the same level as in the case of $\eta=1$. This is due to the additional multiplier enclosed in brackets despite the reduction factor $\eta^2$. In general, the question of the influence of external factors on the output entangled state deserves separate consideration.

\section{TQE from nonlocal photon to two separated odd CV states}

In the previous section we showed that the $TQE$ from a nonlocal photon to two separate $SMSV$ states with maximal output entanglement can be viewed as a probabilistic protocol. In the context of effective distribution of the $CV$ entanglement the chance of success can fall exponentially with further scaling of the entangled network by increasing the number of its members. As it is clear from the previous section the probabilistic nature of the $TQE$ with two input $SMSV$ states is related to the dominance of no-click no-click measurement probability inherent to the $SMSV$ states in which influence of the vacuum state prevails over others.
\par To reduce the impact of no click – no click event, we are going to apply $TQE$ to initially identical odd $CV$ states. Odd $CV$ state can be obtained passing $SMSV$ states through $BS$ with subsequent registration of a single photon in the measuring mode that generates the following heralded $CV$ state (see supplementary material)

\begin{equation}\label{eq:12}
\ket{\Psi_1^{(0)}(y_1)} =
\sqrt{\frac{y_1}{G_1^{(0)}(y_1)}} \sum_{n=0}^{\infty} \frac{y_1^{n}}{\sqrt{(2n+1)!}}\frac{(2n)!}{n!}(2n+1)\ket{2n}, 
\end{equation}

where $G_{1}^{(0)}(y_1)=  \frac{d}{dy_1}(y_1 Z(y_1))=Z^3(y_1)$.

In the modified version of the $TQE$, two initially separate odd $CV$ states interact with a nonlocal photon at two identical $BSs$ with subsequent registration of measurement outcomes $k_1$ and $k_2$ in modes 3 and 4 as shown in \autoref{Figure.2}, (in the figure the $SMSV$ states should be replaced by the ones from which single photon is subtracted). Then, as follows from the supplementary material, the following entangled $CV$ state

\begin{equation}\label{eq:13}
\ket{\triangle^{(1)}_{k_1k_2}(y_2,B)} = \frac{1}{\sqrt{N^{(1)}_{k_1k_2}}}\left(
\begin{array}{c}
\ket{\Psi_{1k_1}^{(00)}(y_2,B)}_1
    \ket{\Psi_{1k_2}^{(01)}(y_2,B)}_2 + \\ 
    b^{(1)}_{k1k2}\ket{\Psi_{1k_1}^{(01)}(y_1,B)}_1
\ket{\Psi_{1k_2}^{(00)}(y_2,B)}_2
\end{array}
\right)
\end{equation}

\!\!\!\!\!\!is formed at the output, where the corresponding CV states that make up the entangled state are presented in the supplementary material.  
\par In general, the entangled state in equation \eqref{eq:13} resembles that presented in the expression \eqref{eq:3} with the exception of the CV components which already have double subscripts and superscripts. So the first subscript 1 and the first superscript 0 indicates the use of the CV states in Eq. \eqref{eq:12}. The second superscript either 0 or 1 is responsible for the parity of the CV state with the same second subscript k showing the number of photons subtracted. The measurement induced CV states originating from the odd CV state in equation \eqref{eq:12} already depends on the parameter $y_2$ which is reduced by $t^2=1⁄(1+B )$ compared to the original $y_1$, that is $y_2=y_1t^2=y_1⁄(1+B)=yt^4= y⁄(1+B)^2$ as shown in the supplementary material. Here we assume that the BS that is used for conditional generation of the CV state $\ket{\Psi_1^{(0)}(y_1)}$ in equation \eqref{eq:12} is identical to the two BSs that are used for measurement induced CV generation in \autoref{Figure.2}. We only note that it is possible to formulate the problem with unequal values of B using different beam splitters, which changes the parameter $y_2$.
\par The CV states with distinct second superscripts, either 0 or 1, and equal k are orthogonal to each other 
$\bra{\Psi_{1k}^{(00)}y_2,B)}\ket{\Psi_{1k}^{(01)}(y_2,B)} = 0$, allowing the entangled state in Eq. \eqref{eq:13} to be considered also in a four-dimensional Hilbert space with the corresponding basic CV states of different parity. The amplitude distortion term $b_{k_1k_2}^{(1)} $ has the same form as in equation \eqref{eq:4}, but with different components. It is presented in the supplementary material. As in the case discussed above, it reduces the degree of the CV entanglement. Moreover, the amplitude distortion factor also determines the normalization factor $N_{k_1 k_2}^{(1)}=1+b_{k_1 k_2}^{(1)2}$.  Only in the case $b_{k_1k_2}^{(1)} =1$ the state in equation \eqref{eq:13} becomes maximally entangled with normalization factor equal to $N_{k_1k_2}^{(1)}=2$. Here, as in the case considered above, the amplitude distorting terms of the interconnected permutations $k_1 \leftrightarrow k_2$ are related to each other as inverse quantities 
$b_{k_1 k_2}^{(1)} = 1 ⁄ b_{k_2 k_1}^{(1)}$.    
\par Using the amplitudes of the hybrid entangled state which are presented in the supplementary material, one can obtain the probability distribution of the coincidence photon counts

\begin{equation}\label{eq:14}
    P_{k_1k_2} = \frac{N_{k1k2}^{(1)}}{2(1+B)Z^{(1)2}(y_1)}
    \begin{cases}
            BZ^{(1)}(y_1)G_{10}^{(01)}(y_2,B), \,\,\text{if}\,\,\, k_1 =k_2 = 0 \\[3ex]
			B\frac{(y_1B)^{k_1}}{k_1!}Z^{(k_1+1)}(y_2)G_{10}^{(01)}(y_2,B), &\!\!\!\!\!\!\!\!\!\!\!\!\!\!\!\!\!\!\! 
            \text{if } k_1>0, k_2 = 0 \\[3ex]
            
			\frac{(y_2B)^{(k_1+k_2-1)}}{k_1!k_2!} k^2_2 Z^{(k_1+1)}(y_2)G_{1k_2}^{(01)}(y_2,B), & \!\!\!\!\text{if } k_1 \ge   0,k_2 > 0 
	\end{cases}
\end{equation}

\!\!\!\!\!\!which can resemble the distribution in Eq. \eqref{eq:5} except that the normalization factor $Z^{(1)}  (y_1)$ of input state is used instead of $\cosh s$
and the normalization factors 
$G_{10}^{(01)}(y_2,B)$ and 
$G_{1k_2}^{(01)}(y_2,B)$ presented in the supplementary material differ from 
$G_{0}^{(1)}(y_1,B)$ and $G_{k_2}^{(1)}(y_1,B)$. 

In the case of identical measurement events $k_1=k_2=k \neq 0$ in two modes, the expression for the probability is simplified 

\begin{equation}\label{eq:15}
    P_{kk}^{(1)}(y_2,B) = 
\frac{(y_2B)^{(2k-1)}k^2Z^{(k+1)}(y_2)G_{1k}^{(01)}(y_2,B)}{(1+B)Z^{(1)2}(y_1)(k!)^2}
\end{equation}

\!\!\!\!\!\!The sum of the probabilities with the same measured number of photons in both modes except no click – no click event can be considered to be total probability 

\begin{equation}\label{eq:16}
    P_{CV}^{(1)} = \sum_{k=1}^{\infty}P_{kk}^{(1)}(y_2,B)
\end{equation}

\!\!\!\!\!for the most effective implementation of the TQE. 
\par The probability of no click – no click event is the product of two probabilities

\begin{equation}\label{eq:17}
    P_{00}^{(1)}(y_2,B) = P_{10}^{(00)}P_{10}^{(01)}
\end{equation}

\!\!\!\!\!where the partial probabilities of no click events in separate measurement channels are given by 

\begin{equation}\label{eq:18}
    \begin{gathered}
        P_{10}^{(00)} = \frac{Z'(y_2)}{Z'(y_1)} = \frac{1}{1+B}\frac{(1-4y_1^2)^{3/2}}{(1-4y_2^2)^{3/2}},\\[3ex]
        P_{10}^{(01)} = \frac{B}{1+B}\frac{\frac{d}{dy_2}(y_2Z'(y_2))}{Z'(y_1)} = 
        2\frac{B}{(1+B)^2}\frac{(1-4y_1^2)^{3/2}(1+2y_2^2)}{(1-4y_2^2)^{5/2}}
    \end{gathered}
\end{equation}

\!\!\!\!\!\!where the subscripts and superscripts used are taken in accordance with those presented in the supplementary material. It is interesting to consider the behavior of the probability $P_{00}^{(1)}$ near $S\approx 0$. By $S\rightarrow 0$ one can get that the maximum value of the probability $P_{00\,max}^{(1)}=1⁄2$ in the case of balanced BS with $B=1$ which is fundamentally different from the case of using two SMSV states for which $P_{00\,max}\approx1$ near $S\approx0$ is observed in the case of using HRBSs with $B\gg1$.
\par If we consider the probability of an event with a coinciding measurement of two photons in separate measurement channels

\begin{equation}\label{eq:19}
    P_{11}^{(1)}(y_2,B) = \frac{(y_2B)Z^{(2)}(y_2)G_{1k}^{(01)}(y_2,B)}{(1+B)Z^{(1)2}(y_1)}
\end{equation}

\!\!\!\!\!\!then we have the following. It can be shown that the probability near $S\approx0$ takes on the next form $P_{11}^{(1)}\approx B(1-B)^2⁄(1+B)^3$  , which has three extreme points. So, in the case of $B=1$, the probability tends to zero $P_{11}^{(1)}(y_2,B=1)\rightarrow{0}$ in the vicinity of $S\approx0$.  A local maximum $P_{11}^{(1)}\approx0.4726$ can be observed in the case of $B=0.2$ near $S\approx0$. More interestingly, the probability can tend to one, i.e. $P_{11}^{(1)}\rightarrow 1$, in the case of using BSs with a sufficiently large value of the parameter B in the vicinity of $S\rightarrow 0$. This estimate shows the usefulness of the application of odd CV states in the TQE protocol. 
\par Numerical results confirm the presented analysis. In \autoref{Figure.5(a,b)}, the B-optimized probability values $P_{CV}^{(1)}$ are shown as a function of the initial squeezing S of the SMSV state. The two curves on the graph represent the optimized finite sum of the probabilities of first four measurement outcomes both with $P_{0-4}=P_{00}^{(1)} +P_{11}^{(1)}+P_{22}^{(1)}+P_{33}^{(1)} +P_{44}^{(1)}$ and without vacuum contribution $P_{1-4}=P_{11}^{(1)}+P_{22}^{(1)}+P_{33}^{(1)} +P_{44}^{(1)}$. Using a limited number of probabilities in summation is possible because the contribution of the probabilities with $k>4$ becomes significantly smaller. As shown in \autoref{Figure.5(a,b)}(a), the probability $P_{0-4}$ is slightly greater than $P_{1-4}$, but the difference between the two probabilities is insignificant $P_{0-4}-P_{1-4}\ll1$ and can be neglected. In general, both $P_{0-4}$ and $P_{1-4}$ can be chosen for the probability of TQE to two odd CV states, i.e., either $P_{CV}^{(1)}=P_{0-4}$ or $P_{CV}^{(1)}=P_{1-4}$. Vacuum fluctuations manifest themselves as temperature changes in the energy density, leading to tiny fluctuations in heat in the TES detector. Regardless of the accepted estimate, either $P_{CV}^{(1)}=P_{0-4}$ or $P_{CV}^{(1)}=P_{1-4}$, the probability of the TQE to odd CV states takes on significant values $P_{CV}^{(1)}>0.98$ at any value of initial squeezing S which only slightly decreases with increasing the S.  
\par As for the optimizing values of the beam splitter parameter B, it turns out that they are equal to each other $B_{CV}^{(1)}=B_{0-4}=B_{1-4}=276.6$ and do not depend on S, as shown in \autoref{Figure.5(a,b)}(b). Since the values coincide, they are shown in \autoref{Figure.5(a,b)}(b) as one horizontal line. Regardless of the selected probability in \autoref{Figure.5(a,b)}(a), either $P_{0-4}$ or $P_{1-4}$, the beam splitter parameter B that provides them is the same $B_{CV}^{(1)}$ in \autoref{Figure.5(a,b)}(b).    
\par From the given numerical values it is clear that $B_{CV}\gg B_{CV}^{(1)}$, but nevertheless, the $B_{CV}^{(1)}$ obtained from \autoref{Figure.5(a,b)}(b) can also reduce the initial value of the squeezing parameter $y_2=y_1⁄(1+B_{CV}^{(1)})$, which can lead to a decrease in the average number of photons of the CV states. To increase the brightness of the output CV entanglement, highly quadrature squeezed SMSV states with squeezing $S>10\,\,dB$ can be used. A practical way to improve efficiency of the TQE protocol with maximum output CV entanglement is to reduce $B_{CV}^{(1)}$. Indeed, $B_{CV}^{(1)}$ is a parameter whose reduction can increase the output brightness of the CV states from which CV entanglement is formed. So if we reduce the value of $B_{CV}^{(1)}$ to 10, that is, $B_{CV}^{(1)} =10$, then the overall probability of the measurement induced maximal CV entanglement turns out to be $P_{CV}^{(1)}>0.968$ for all those values S which are used in \autoref{Figure.5(a,b)}(a). It is possible to further reduce the value of $B_{CV}^{(1)}<10$ while keeping the probability of maximal TQE at a sufficiently high level and maintaining the output brightness of the CV entanglement. Since increasing the transmittance of the beam splitters used is practically feasible, the importance of the TQE from a nonlocal photon to two initially separated odd CV states only increases.

\begin{figure}[htbp]
\centering\includegraphics[width=0.9\textwidth]{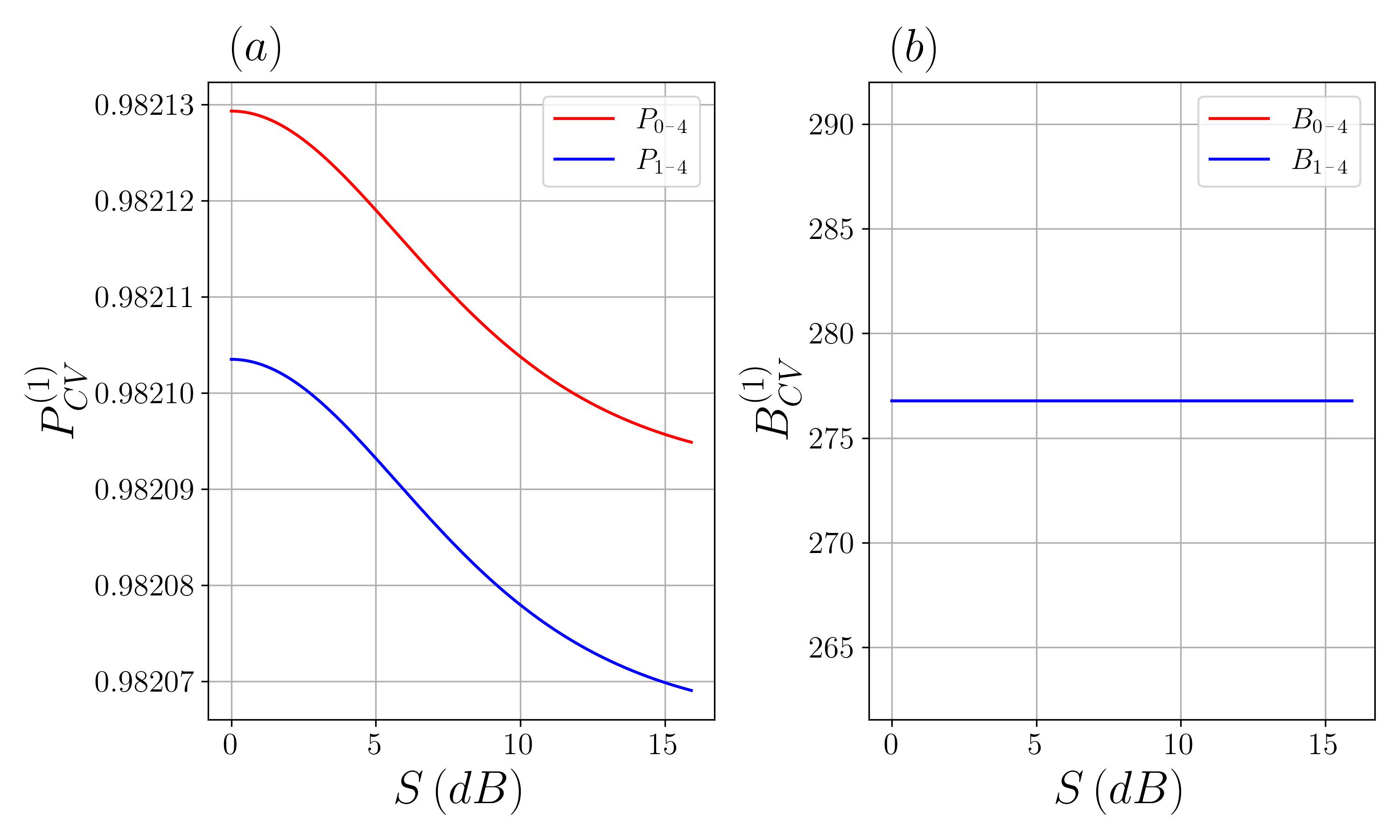}
\caption{(a,b) (a) Plots of the TQE probabilities $P_{CV}^{(1)}$, optimized over $B=B_{CV}^{(1)}$, versus the initial squeezing $S$ when using the input CV states in equation \eqref{eq:12}. Two dependencies correspond to cases with inclusion of the vacuum contribution $P_{0-4}$ and without it $P_{1-4}$. Including the vacuum contribution only slightly increases the overall probability $P_{CV}^{(1)}$ of maximum output CV entanglement. The optimizing values $B_{0-4}$ and $B_{0-1}$ coincide in (b), i.e. $B_{CV}^{(1)} =B_{0-4}=B_{1-4}=276.6$, which is expressed by one horizontal line.}
\label{Figure.5(a,b)}
\end{figure}

\section{Conclusion}
A nonlocal photon propagating simultaneously in two modes can carry maximum entanglement. Moreover, such entanglement is quite easy to create in the presence of a single photon by passing it through the balanced beam splitter. It is known that nonlocality of the single photon entangled states is not revealed in a standard Bell test with only linear optical elements \cite{14}. Nevertheless, the entanglement of a single photon can be converted to an entangled state consisting of CV states of definite parity. Here, we have examined in detail the mechanism of entanglement transfer from a nonlocal photon to two initial CV states of a certain parity, in which there is no direct CV-CV interaction. As the CV states, two cases are considered: SMSV states and odd CV states, i.e. SMSV states from which one photon is initially subtracted. The CV states can be located at a certain distance from each other, and the generated CV entanglement can be of a distributed nature which partially brings this approach closer to the entanglement swapping protocol \cite{39}. But in the TQE protocol initially separated CV states interact only with the nonlocal photon through a system of two identical BSs and the initial CV entanglement is absent as shown in \autoref{Figure.2}(c). Moreover, the protocol is free from BSM.
\par The TQE on demand occurs when photon number measurements are observed in the measurement modes. In addition to the emergence of the CV entanglement, there is a transformation of initially Gaussian states into non-Gaussian ones that retains their parity. The TQE cannot occur ideally, i.e. in such a way that a nonlocal balanced photon is automatically converted to a maximally parity-entangled CV state due to the appearance of an amplitude distortion factor, as if the state had additionally passed through a noisy channel. Note that the SCS transforms into an entangled coherent state after passing through balanced beam splitter. The difficulties in quantum engineering of the SCSs, as well as the nonorthogonality of coherent states with amplitudes equal in magnitude but opposite in sign, are well known \cite{23}\cite{24}\cite{25}\cite{26}\cite{27}.
\par What is surprising is that the measurement result of vacuum-two photons and two photons-vacuum turns out to be the best choice for the TQE from nonlocal photon to two separate SMSV states. The probability of the outcomes limited by value 0.2344 is higher than the sum of the probabilities of the measurement results with the same number of photons in both measurement channels. Moreover, such a realization can be implemented with almost balanced BSs which allows to maintain the brightness of the CV states forming the CV entanglement at the proper level. Such a TQE from nonlocal photon to two SMSV states can be considered perfect for the SMSV states, since the use of HRBSs is not practically appropriate. Despite the relevance of the scheme for entanglement distribution, its probabilistic nature imposes substantial practical limitations. 
\par To significantly improve the functionality of our protocol and its potential in developing quantum technologies, we have considered TQE from nonlocal photon to two separate odd non-Gaussian states. In contrast to the case considered above, the sum of the probabilities of simultaneous measurement of the same number of photons in both measuring channels is maximum for the CV states. Such on-demand transfer of quantum entanglement from nonlocal photon to odd CV states can be achieved with probability close to unity but with a reduction in the brightness of the output non-Gaussian states due to the use of a BS with a higher reflectivity. Reducing the beam splitting parameter allows us to find a compromise between the output probability and the average number of photons in CV states that form entanglement. Based on the above analysis, it should be recognized that the TQE from nonlocal photon to two odd CV states is perfect, since transfer of maximum entanglement can be realized with high probability and with preservation of a sufficiently large average number of photons.
\par In conclusion, we have demonstrated that adding a nonlocal photon to both Gaussian and non-Gaussian states allows precise control over the design of the output CV entanglement. The essence of our work is to demonstrate the design of CV entanglement at a new fundamental level, where the probability, degree of entanglement and average number of photons in quantum states are controlled by the initial conditions. It turned out that the efficiency of the TQE protocol largely depends on the CV states into which the entanglement is transferred. Overall, we believe that the possibility of controlling TQE from nonlocal photon to the CV states opens the way to new quantum protocols in the mesoscopic regime, which is of particular importance for quantum metrology and construction of quantum networks.

\begin{backmatter}

\bmsection{Funding}
\!\!\!The study was supported by the grant of the Russian Science Foundation No. 25-12-20026, https://rscf.ru/project/25-12-20026/.

\bmsection{Acknowledgment}
 \!\!\!\!\!\!Supplementary material for the manuscript was prepared with the support of the Foundation for the Advancement of Theoretical Physics and Mathematics “BASIS” (Project № 24-1-1-87-1). 

\bmsection{Disclosures}
\!\!\!The authors declare no conflicts of interest.

\bmsection{Supplemental document}
\!\!\!See \href{https://opg.optica.org/josab/abstract.cfm?uri=josab-43-3-433#articleSupplMat}{supplementary material} for supporting content.

\end{backmatter}

\end{document}